\documentclass[aps,superscriptaddress,prl,preprint]{revtex4}
\usepackage{graphicx}
\begin{document}
\title{Nonlocal transport in the charge density waves of
$o$-TaS$_3$}
\author{Katsuhiko Inagaki}
\affiliation{Division of Applied Physics, Graduate School of Engineering,
Hokkaido University, Kita 13 Nishi 8 Kita-ku, Sapporo, 060-8628, Japan}
\affiliation{Center of Education and Research for Topological Science and Technology,
Hokkaido University, Kita 13 Nishi 8 Kita-ku, Sapporo, 060-8628, Japan}
\author{Masakatsu Tsubota}
\affiliation{Division of Applied Physics, Graduate School of Engineering,
Hokkaido University, Kita 13 Nishi 8 Kita-ku, Sapporo, 060-8628, Japan}
\author{Satoshi Tanda}
\affiliation{Division of Applied Physics, Graduate School of Engineering,
Hokkaido University, Kita 13 Nishi 8 Kita-ku, Sapporo, 060-8628, Japan}
\affiliation{Center of Education and Research for Topological Science and Technology,
Hokkaido University, Kita 13 Nishi 8 Kita-ku, Sapporo, 060-8628, Japan}
\begin{abstract}
We studied the nonlocal transport of a quasi-one dimensional conductor $o$-TaS$_3$.
Electric transport phenomena in charge density waves include
the thermally-excited quasiparticles, and collective motion of charge density waves (CDW).
In spite of its long-range correlation, the collective motion of a CDW
does not extend far beyond the electrodes, where phase slippage breaks
the correlation. We found that nonlocal voltages appeared in the CDW of $o$-TaS$_3$,
both below and above the threshold field for CDW sliding.
The temperature dependence of the nonlocal voltage
suggests that the observed nonlocal voltage
originates from the CDW even below the threshold field. Moreover, our observation
of nonlocal voltages in both the pinned and sliding states
reveals the existence of a carrier with long-range correlation,
in addition to sliding CDWs and thermally-excited quasiparticles.
\end{abstract}
\maketitle
Nonlocal properties in charge density waves (CDW) have been attracting interest for decades \cite{Gill1982,Ramakrishna1992,Lemay1996}.
Since electric transport phenomena include 
collective motion \cite{Gruner},
it is natural to expect a sliding CDW to be a possible source of 
nonlocal transport \cite{Gill1982}. 
The correlation of the CDW phase has been estimated in the 1 to 100 $\mu$m range \cite{Gill1982,Mihaly1983,Thorne1985,Lyding1986,Borodin1987,Sweetland1990,McCarten1992,Danneau2002,LeBolloch2008,Tsubota2009},
which is long enough to allow experimental studies. 
The correlation inside the sliding region has been confirmed.
However, it has been shown that
the correlation of a sliding CDW does not extend far beyond the electrodes,
where phase slippage may destroy the correlation \cite{Ramakrishna1992,Lemay1996}. 

\begin{figure}
\includegraphics[scale=0.6]{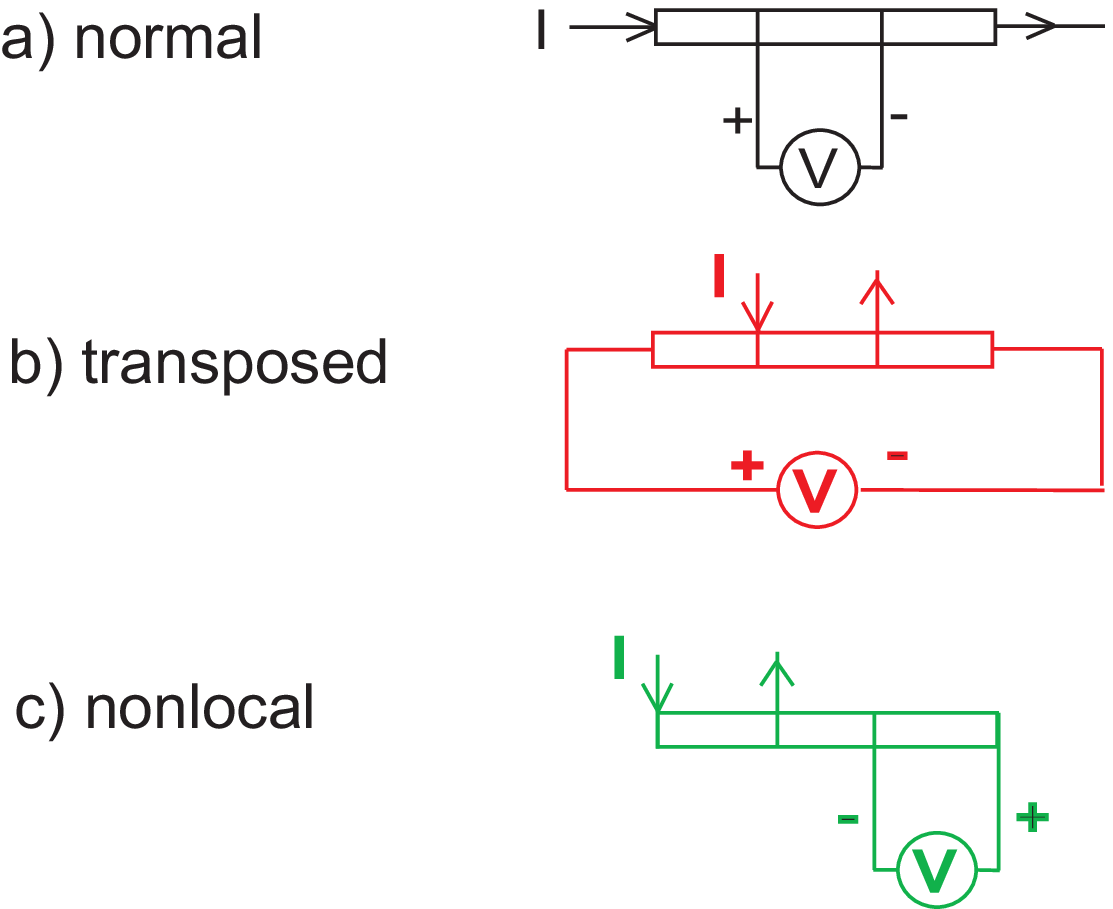}
\caption{Configurations for the measurements: a) normal configuration, 
b) transposed configuration, and c) nonlocal configuration. The polarity of the voltmeter in the nonlocal configuration provides a positive value when the spreading resistance is dominant. }\label{fig1}
\end{figure}

In this article, we report a nonlocal transport phenomenon with a 
quasi-one dimensional conductor $o$-TaS$_3$.
Following the pioneering work by Gill \cite{Gill1982},
we compared current-voltage characteristics obtained with normal and transposed  electrode configurations, as well as with nonlocal detection (Fig. \ref{fig1}). 
We found that nonlocal voltages appeared in the CDW of $o$-TaS$_3$,
both below and above the threshold field for CDW sliding.
The sign of the observed nonlocal voltage  was opposite 
to that of the spreading resistance \cite{Borodin1986,Slot2001}. In addition to this,
the temperature dependence of the nonlocal voltage
suggests that the observed nonlocal voltage
originates from the CDW. Moreover, our observation
of nonlocal voltages in both the pinned and sliding states
reveals the existence of a carrier with long-range correlation,
in addition to sliding CDWs and thermally-excited quasiparticles.
Finally, we compared our observation to the coexistence of two CDWs in $o$-TaS$_3$ \cite{Inagaki2008,Toda2009}, and found it is
closely related to topological dislocations in CDWs.

We performed a four-probe measurement,
which requires a voltmeter with a high input impedance so that the leakage current to the voltmeter becomes negligible.
For fully-gapped CDW materials, such as $o$-TaS$_3$, the use of four-probe measurement is usually
difficult especially at low temperatures where the sample resistance
often exceeds the typical input impedance of a voltmeter, for example,
$Z_{in} \sim 1$ G$\Omega$. 
We exploited an electrometer (Keithley 6512, $Z_{in} > 200$ T$\Omega$)
to measure the voltage drop of the $o$-TaS$_3$ samples.
 
$o$-TaS$_3$ crystals were synthesized with a standard chemical vapor transport
method.
A pure tantalum sheet and sulfur powder were put in a
quartz tube. The quartz tube was evacuated to $1 \times 10^{-6}$ Torr
and heated in a furnace at 530 ${}^\circ$C for two weeks. 
The electrodes were made using 50-$\mu$m-diameter silver wires glued with
silver paint. Gold thin film was deposited on the crystal before
the silver wires were attached to reduce the contact resistance
to 1 $\Omega$ at room temperature.
The spacing between the electrodes was typically 200 $\mu$m,
and the sample cross section was $50 \times 5$ $\mu$m$^2$ (width $\times$ thickness).
 
\begin{figure}
\includegraphics[scale=0.6]{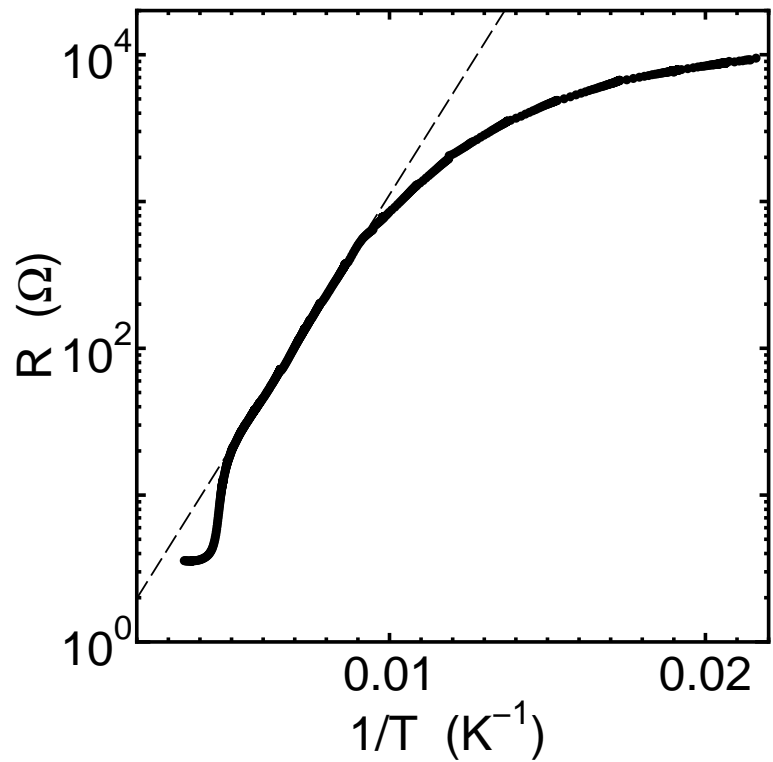}
\caption{Temperature dependence of sample resistance.
The broken line is an Arrhenius fit $R \propto \exp (T/T_0)$ with $T_0 = 790$ K.}\label{fig2}
\end{figure}

Figure \ref{fig2} shows the temperature dependence of the resistance
of a $o$-TaS$_3$ crystal. 
The result shows the well-known features of $o$-TaS$_3$,
a Peierls transition temperature of 220 K, an Arrhenius-type temperature
dependence with an activation energy of 790 K, and
a deviation from the Arrhenius plot below
100 K. We checked the contact resistance by comparing
the four- and two-terminal resistances,
and found it was negligible in the room temperature to 40 K range.
We also confirmed that the result was unaffected by
the difference between substrates by comparing
data obtained with the silica and sapphire
substrates. Based on these considerations, we confirmed that
the four-terminal measurement was properly performed
over the entire temperature range of the measurement.

\begin{figure}
\includegraphics[scale=0.6]{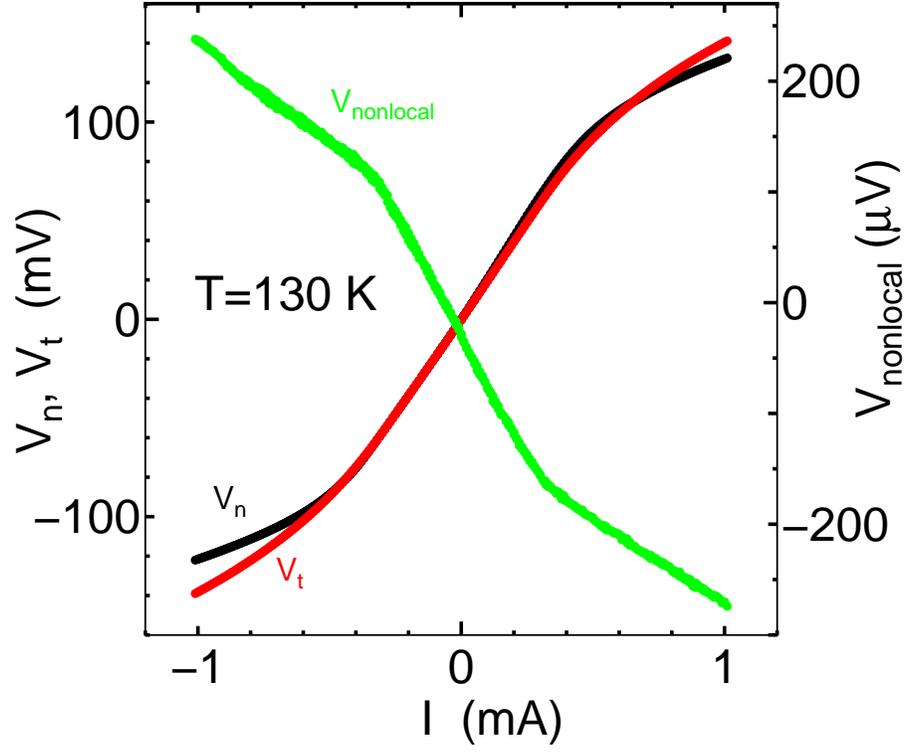}
\caption{Current-votlage characteristics at 130 K for the
normal, transposed, and nonlocal configurations, denoted by
black, red, and green curves, respectively.}\label{fig3}
\end{figure}

Figure \ref{fig3} shows the current-voltage ($I$-$V$) characteristics
at 130 K in the normal and transport configurations, as well as the
nonlocal voltage.
The $I$-$V$ characteristics with the normal configuration
(black line) and with the transposed configuration (red line)
coincide in a low current regime where ohmic behavior is
observed. Above the threshold of nonlinear conduction,
we detected a discrepancy between the two configurations.
This discrepancy can be understood as the phase-slip voltage
accompanied by CDW sliding \cite{Gill1982,Ramakrishna1992}.
When the CDW slides under a condition where the regions outside the current electrodes
are strongly pinned, elastic deformation is induced in the CDW.
In the vicinity of each electrode, the elastic
deformation relaxes by breaking CDW phase continuity.
In the transposed configuration (Fig. \ref{fig1} b), 
the measured voltage is affected by the phase slip, in contrast
to the normal configuration (Fig. \ref{fig1} a), hence
the difference between these two configurations as shown in Fig. \ref{fig3}
reveals that
the phase slip really occurred in the sample.

We also observed nonlocal voltage as shown in Fig. \ref{fig3} (green line),
by using the nonlocal configuration (Fig. \ref{fig1} c).
The result shows that at 130 K, a negative voltage was detected at a pair of electrodes 200 $\mu$m away.
The magnitude of the nonlocal voltage was around 200 $\mu$V at 1 mA.
This voltage was large enough to be measured with our ultra-high impedance
voltmeter, whose minimum detectable voltage was 10 $\mu$V.
The voltage was proportional to
the injected current, with kinks at $\pm 0.3$ mA, above which the slope of
the $I$-$V$ curve became moderate.
The emergence of a voltage at the nonlocal probes is fascinating since
Gill reported that there was no
detectable signal in the CDW state of NbSe$_3$ \cite{Gill1982}.
Such an attempt has also been made for other CDW systems, however, the nonlocal effects
were unlikely to extend over the electrode, except via strain distribution \cite{Lemay1996}.

We first compared our observation with a trivial source,
known as spreading resistance, which could affect
the nonlocal voltage probes \cite{Borodin1986,Slot2001}. 
When the current is injected from a tiny electrode into a bulk sample,
it cannot flow homogeneously until it spreads over a cross section. 
In the vicinity of the electrodes the flow lines are curved, and 
some of them travel in the opposite direction to the counterelectrode.
At the sample surface the electric potential $U(x)$ of position $x$ is expressed by
\begin{equation}
U(x)=-E\frac{t'}{\pi}\mathrm{arc cosh}
\left|
\frac{\cosh \left({\pi l}/{t'}\right)
-\exp \left( {\pi x}/{t' }\right) }
{\sinh \left({\pi l}/{2t' } \right) }
\right|,\label{eq1}
\end{equation}
where $E$ denotes an applied field, $l$ is the electrode length, and $t'=t\sqrt{A}$ is a reduced thickness with $A$ being 
the anisotropic ratio of conductivity \cite{Borodin1986,Slot2001}.
In addition, the effective cross section for the current is limited
to the specific distance $\sim t'$, and an additional contribution to resistance
emerges, known as spreading resistance. An important
feature of spreading resistance is that it obeys Ohm's law
since its nature is the restriction of the current path.
In our configuration for nonlocal detection (Fig. \ref{fig1} c), the contribution of the spreading resistance 
must generate a \textit{positive} voltage between the voltage probes.

\begin{figure}
\includegraphics[scale=0.6]{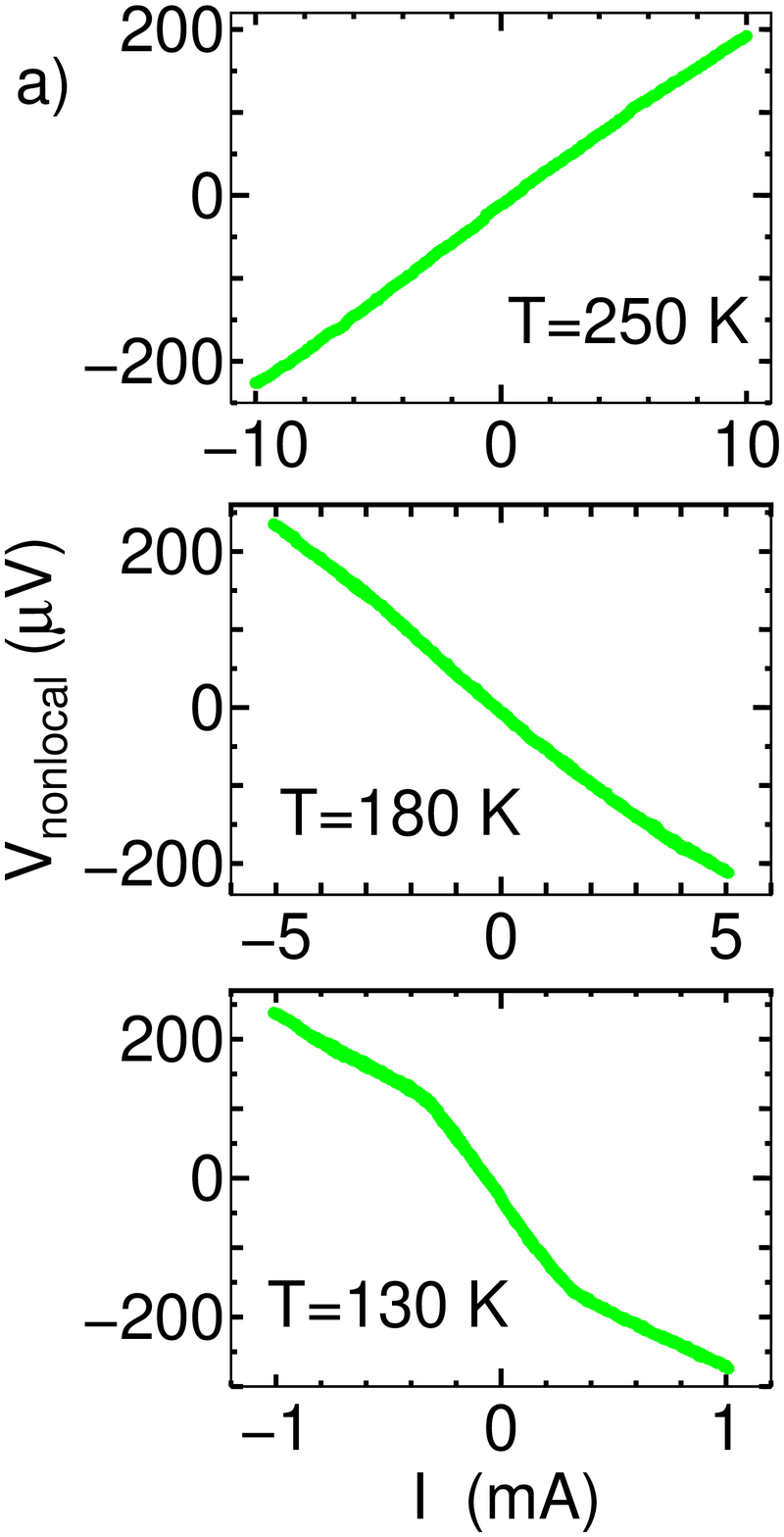}
\hspace{2ex}
\raise3cm\hbox{\includegraphics[scale=0.4]{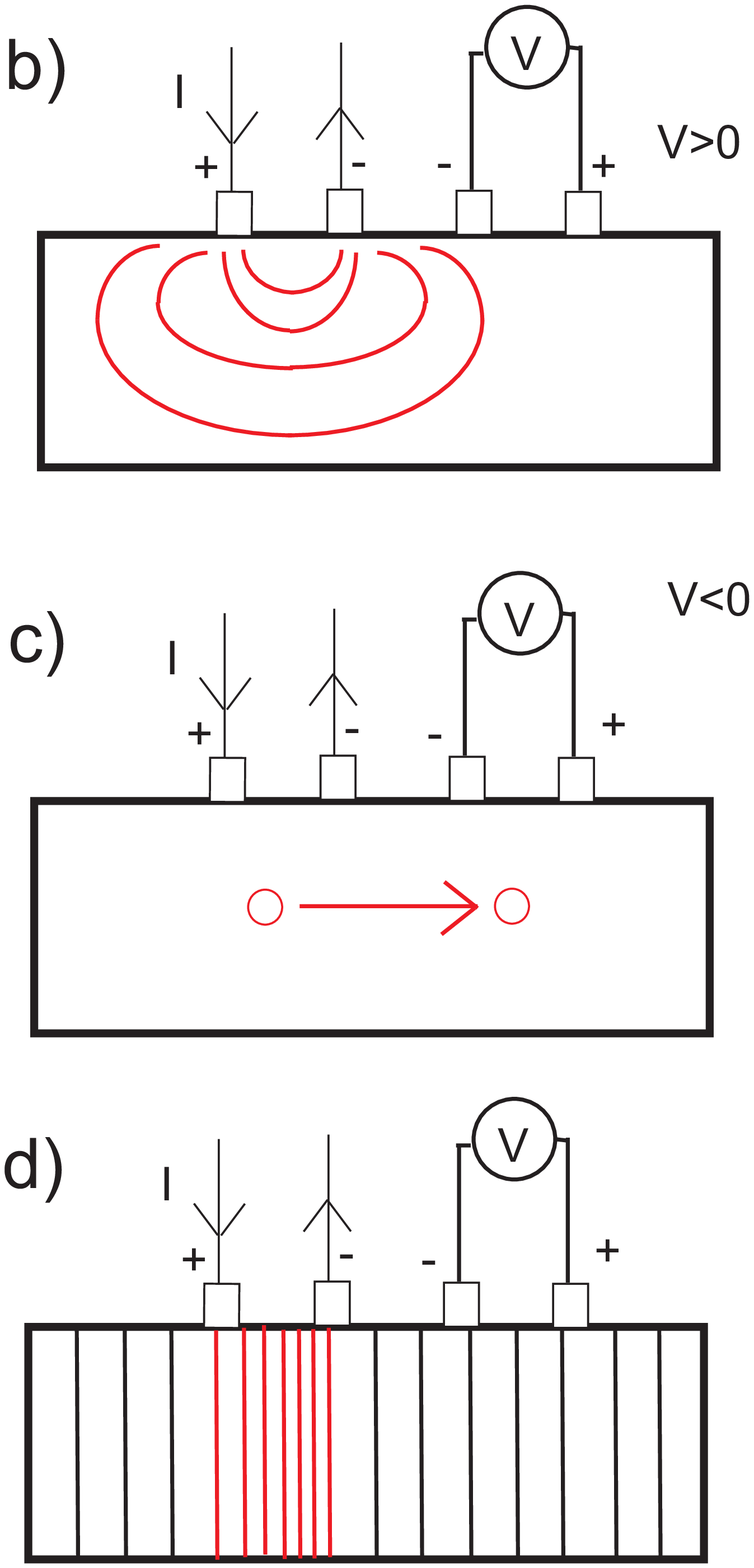}}
\caption{a) Temperature evolution of the nonlocal voltages at 130, 180, and
250 K. Polarity inversion was seen below and above the Peierls transtion,
220 K. b) Schematic diagram of flow lines to provide spreading resistance. c) Possible correlation between current and voltage probes via dislocation lattice. d) Correlation of sliding CDW is fig1 exceeded over current probes.}\label{fig4}
\end{figure}

In fact, the observed nonlocal voltages were \textit{negative}
as shown in Fig. \ref{fig3}. 
Moreover, the temperature dependence of the nonlocal voltage
suggests that CDW is closely related to its origin.
Figure \ref{fig4}a shows the temperature evolution of
the nonlocal voltages at 130, 180, and 250 K.
At 180 K, although the kink and nonlinearity became less pronounced,
the nonlocal voltage was negative.
In contrast, at 250 K, above the Peierls transition ($T_P = 220$ K), 
the polarity was reversed, and became positive.
The slope of the $I$-$V$ curve was $2.1 \times 10^{-2}$ $\Omega$.
This is in good agreement with the estimation of spreading resistance $2.8 \times 10^{-2}$
$\Omega$ estimated using Eq. (\ref{eq1}) with an anistropy ratio of $A=500$ \cite{Slot2001}.
To confirm this is not an artifact from the measurement circuit,
we performed several sets of experiments, including 
a comparison of the data while inverting the polarity of the electrodes
of both the current generator and the voltmeter. We also checked
the reproducibility by using different specimens and cryostats.
All the data were consistent and we concluded that the 
observed \textit{negative} nonlocal voltage was really a
property of the samples. 
Hence it is unlikely that 
the nonlocal voltage can be attributed to the trivial spreading resistance,
whereas its attribution to CDW is plausible.

In addition, 
the voltages were nonlinear functions of the current with
apparent kinks (Fig. \ref{fig3}).
The positions of the kinks in the nonlocal voltage were found
to be close to the threshold of nonlinear conduction of the CDW.
When the sample was in a CDW state the Peierls gap opens at Fermi surfaces, 
so that thermally-activated quasiparticles
and collective motion of the CDW contribute to electric conduction.
The former provides ohmic currents, and as discussed above,
might contribute to positive nonlocal voltages.
Hence, a quasiparticle is also ruled out as a major contributor
to the nonlocal voltage,
and the collective motion becomes a candidate for the nonlocal voltage.
Nevertheless, the observed kinks are clearly related to
the onset of CDW sliding.

For the sliding CDW to provide the nonlocal voltage,
probed 200 $\mu$m away from the emitting electrodes,
the phase correlation length of the CDW should be at least
comparable to or longer than the travelling length.
There are several estimations for the phase correlation
length, ranging from 1 to 100 $\mu$m \cite{Gill1982,Mihaly1983,Thorne1985,Lyding1986,Borodin1987,Sweetland1990,McCarten1992,Danneau2002,LeBolloch2008,Tsubota2009}. X-ray diffraction
typically provides the lengths of the order of 1 $\mu$m \cite{Sweetland1990,Danneau2002,LeBolloch2008},
whereas other kinds of estimations based on CDW dynamics
provide longer values of up to 100 $\mu$m \cite{Gill1982,Thorne1985,Lyding1986,Tsubota2009}. 
Moreover, the phase
coherence length is a strong function of defect concentration, and
will differ for different materials and even for samples of the same
material.
Consequently,
a consideration of the correlation length does not deny
the possibility of a sliding CDW being the origin
of the nonlocal voltage.

However, even if the CDW correlation
could extend such a distance, 
the standard interpretation of the phase-slip voltage
suggests that the phase correlation might be lost 
in the vicinity of the electrode \cite{Gill1982,Ramakrishna1992}.
As shown in Fig. \ref{fig3}, the difference between the voltages of the normal and transposed
configurations implies that the CDW is in the sliding state with
a considerable rate of phase slip \cite{Ramakrishna1992}. 
Moreover, it should be noted that the nonlocal voltage was observed even
below the kinks. Hence, the source of the nonlocal voltage
must be present whether or not the CDW slides.

Now we may propose a possible model based on the
above discussions.
Since the CDW phase correlation is destroyed in the vicinity of
the electrodes, and the nonlocal voltage is also seen below the
kinks,  \textit {we must seek for a different transport mechanism originated
from the CDW state}, rather than sliding.
Its properties should include ohmic behavior, and 
it should possess a spatial correlation over at least 200 $\mu$m. 
A recent synchrotron diffraction study reports that two kinds of
CDWs, incommensurate and commensurate CDWs, coexist in
$o$-TaS$_3$ \cite{Inagaki2008}, and a subsequent
ultrafast optical pump-probe study of $o$-TaS$_3$ also provides
evidence for two types of CDW phases \cite{Toda2009}.
Discommensuration,
or a lattice of topological dislocations \cite{McMillan1976} exists from the Peierls temperature
to around 50 K \cite{Inagaki2008}. Each topological dislocation is accompanied
by a charge, 
which can move along the chain direction of a CDW.
The distance between dislocations is determined by the difference
between the CDW wavelength and the lattice constant of the pristine lattice.
The advantage of this model as regards explaining the nonlocal voltage
relies on the correlation length of the dislocation lattice.
Hence the question now moves to how far can the correlation be extended. 

Coulomb repulsion plays a major role in the interaction between the dislocations
since each dislocation is accompanied by an electric charge.
In some CDW materials, such as NbSe$_3$, uncondensed
electrons remain at the Fermi surface, leading to the screening of the Coulomb potential.
In the case of $o$-TaS$_3$, the entire Fermi surface contributes to the CDW,
and at low temperatures the density of thermally-activated quasiparticles is 
very low, so that the screening of the Coulomb potential becomes negligible.
This helps to increase the correlation length of the dislocation lattice.
We believe this is the most plausible model for explaining the observed nonlocal
voltage.

Consequently, the observed nonlocal voltage may have three 
difference origins, as expressed by
\begin{equation}
V_\mathrm{nonlocal} = V_\mathrm{spread} + V_\mathrm{dislocation} + V_\mathrm{sliding},\label{eq2}
\end{equation}
where $V_\mathrm{spread}$,  $V_\mathrm{dislocation}$, and $V_\mathrm{sliding}$
denote spreading resistance, the correlated motion of the dislocation
lattice, and CDW sliding, respectively. Under our experimental conditions,
each term contributes as follows: $V_\mathrm{spread} > 0$ for all temperatures;
$V_\mathrm{dislocation} <0$ and $|V_\mathrm{spread}|<|V_\mathrm{dislocation}|$ below the Peierls temperature; and
$V_\mathrm{sliding}>0 $ above the CDW sliding threshold. Each term is
schematically illustrated in Figs. \ref{fig4}b-\ref{fig4}d. By using these terms,
Eq. (\ref{eq2}) provides
qualitative explanations for the observed peculiar behavior as shown in Figs. \ref{fig3}
and \ref{fig4}. It should be noted that the polarity of $V_\mathrm{sliding}$
is positive, thus coinciding with $V_\mathrm{spread}$. This might be caused by the
backflow of quasiparticle current accompanied by CDW sliding \cite{vdZ2001}.

Finally, we discuss the possible relationship between our observation and 
the soliton transport in a commensurate CDW \cite{Gruner}.
Previous studies have suggested that the low temperature transport of $o$-TaS$_3$
should be understood as a CDW soliton \cite{Maki1977,Takoshima1980}. Experimentally, the anisotropy of conductivity was
found to be increased at low temperatures along the chain direction \cite{Takoshima1980}.
Although the excitation energy was estimated to be 250 K, 
the previous results failed to detect any 
traces of the soliton at temperatures at which they should be excited.
Assume that our observation of nonlocal voltage is an emergence of soliton transport. The phase diagram of $o$-TaS$_3$ \cite{Inagaki2008} shows that there is commensurate CDW at 130 K. Hence, if the life time of the excited soliton is long
enough to travel over the 200 $\mu$m, it can be a candidate to the origin
of the nonlocal voltage. 
Since we lack unambiguous evidence for the relationship to the CDW soliton
within our study, we will pursue the nature of this nonlocal voltage 
in further studies, including the use of optical \cite{Toda2009}, acoustic \cite{Malomed1993} and diffraction \cite{Inagaki2008,LeBolloch2008} methods.

To summarize, 
we studied the nonlocal transport of a quasi-one dimensional conductor $o$-TaS$_3$.
We found that nonlocal voltages appeared in the CDW of $o$-TaS$_3$,
both below and above the threshold field of CDW sliding.
The temperature evolution of the nonlocal voltage
suggests that the observed nonlocal voltage
originates from the CDW. Moreover, our observation
of nonlocal voltage in both the pinned and sliding states
reveals the existence of a carrier with long-range correlation,
in addition to a sliding CDW and thermally-excited quasiparticles.

We are grateful to K. Ichimura, K. Yamaya, Y. Toda, M. Hayashi and Y. Nogami for fruitful discussions.
One of the authors (K.I.) thanks the Japan Society for the Promotion of Science
for financial support via a Grand-in-aid for Scientific Research (Kakenhi).

\clearpage

\end{document}